\documentclass[a4paper,12pt]{article}

\usepackage{amsfonts,amssymb,cite}  
\usepackage[nointlimits]{amsmath}

\numberwithin{equation}{section}
\numberwithin{theorem}{section}

\newcommand{\mc}[1]{{\mathcal #1}}

\newcommand{\bb}[1]{{\mathbb #1}}

\newcommand{\id}{{1 \mskip -5mu {\rm I}}}
\renewcommand{\epsilon}{\varepsilon}

\begin{document}

\begin{titlepage}
\par\vskip 1cm\vskip 2em

\begin{center}
{\Large  \textbf{Towards a nonequilibrium thermodynamics:
a self-contained macroscopic description of driven diffusive systems} }

\bigskip

\begin{tabular}[t]{c}
$\mbox{L. Bertini}^{1}\!\!\phantom{m}\mbox{ A. De Sole}^{2}
\!\phantom{m}\mbox{D. Gabrielli}^{3}\!\phantom{m}
\mbox{G. Jona--Lasinio}^{4}\!\phantom{m}\mbox{C. Landim}^{5}$
\\
\end{tabular}
\par
\medskip
{\footnotesize
\begin{tabular}[t]{ll}
{\bf 1} &
Dipartimento di Matematica, Universit\`a di Roma La Sapienza\\
&  P.le A.\ Moro 2, 00185 Roma, Italy\\
&  E--mail: {\tt bertini@mat.uniroma1.it}\\
{\bf 2} & Dipartimento di Matematica, Universit\`a di Roma La Sapienza, Roma, Italy\\
&  Mathematics Department, Harvard University, Cambridge MA, USA\\
&  E--mail: {\tt desole@mat.uniroma1.it}\\
{\bf 3} & Dipartimento di Matematica, Universit\`a dell'Aquila\\
&  67100 Coppito, L'Aquila, Italy \\
&  E--mail: {\tt gabriell@univaq.it}\\
{\bf 4} & Dipartimento di Fisica and INFN, Universit\`a di Roma La Sapienza\\
&  P.le A.\ Moro 2, 00185 Roma, Italy\\
&  E--mail: {\tt gianni.jona@roma1.infn.it}\\
{\bf 5}& IMPA, Estrada Dona Castorina 110, J. Botanico, 22460 Rio
de Janeiro, Brazil\\
& CNRS UPRES--A 6085, Universit\'e de Rouen, \\
&
76128 Mont--Saint--Aignan Cedex, France \\
& E--mail: {\tt landim@impa.br}\\
\end{tabular}
}
\end{center}

\begin{abstract}
\noindent
In this paper we present a self-contained macroscopic description of
diffusive systems interacting with boundary reservoirs and under the
action of external fields.  The approach is based on simple postulates
which are suggested by a wide class of microscopic stochastic models
where they are satisfied.  The description however does not refer in
any way to an underlying microscopic dynamics: the only input required
are transport coefficients as functions of thermodynamic variables,
which are experimentally accessible.  The basic postulates are local
equilibrium which allows a hydrodynamic description of the evolution, 
the Einstein relation among the transport coefficients, 
and a variational principle defining the out of equilibrium free
energy.  Associated to the variational principle there is a
Hamilton-Jacobi equation satisfied by the free energy, very useful for
concrete calculations.  Correlations over a macroscopic scale are, in
our scheme, a generic property of nonequilibrium states. Correlation
functions of any order can be calculated from the free energy
functional which is generically a non local functional of
thermodynamic variables. Special attention is given to the notion of
equilibrium state from the standpoint of nonequilibrium.
\end{abstract}

\vfill

\end{titlepage}
\vfill\eject


\section{Introduction}
\label{s:1}

The main purpose of the present paper is to describe a ``physical
theory" for a certain class of thermodynamic systems out of
equilibrium, which is founded on and supported by the analysis of a
large family of stochastic microscopic models.  Out of equilibrium the
variety of phenomena one can conceive makes it difficult to define
general classes of phenomena for which a unified study is possible.
Furthermore the details of the microscopic dynamics play a far greater
role than in equilibrium.  Since the first attempts to construct a non
equilibrium thermodynamics, a guiding idea has been that of {\it local
  equilibrium}, which means that locally on the macroscopic scale it
is possible to define thermodynamic variables like density,
temperature, chemical potentials... which vary smoothly on the same
scale. Microscopically this implies that the system reaches local
equilibrium in a time which is short compared to the times typical of
macroscopic evolutions, as described for example by {\it hydrodynamic
  equations}.  There are important cases however where local
equilibrium apparently fails like aging phenomena in disordered
systems due to insufficient ergodicity. These will not be considered
in this paper. Also the case in which magnetic fields play a role is
not covered by our analysis.

The simplest nonequilibrium states one can imagine are {\it stationary
states} of systems in contact with different reservoirs and/or under
the action of external (electric) fields. In such cases, contrary to equilibrium,
there are currents (electrical, heat, matter of various chemical constitutions
...) through the system whose macroscopic behavior is encoded in transport
coefficients like the diffusion coefficient, the conductivity or the mobility.

The ideal would be to approach the study of these states starting from
a microscopic dynamics of molecules interacting with realistic forces and
evolving with Newtonian dynamics. This is beyond the reach of present
day mathematical tools and much simpler models have to be adopted in
the reasonable hope that some essential features are adequately
captured.  In the last decades stochastic models of interacting
particle systems have provided a very useful laboratory for studying
properties of stationary nonequilibrium states.  From the study of
these models has emerged a macroscopic theory for nonequilibrium
diffusive systems which can be used as a phenomenological theory.

A basic issue is the definition of nonequilibrium thermodynamic
functions.  For stochastic lattice gases a natural solution to this
problem has been given via a theory of dynamic large deviations
(deviations from hydrodynamic trajectories) and an associated
variational principle leading to a definition of the free energy in
terms of transport coefficients.

One of the main differences between equilibrium and nonequilibrium
systems, is that out of equilibrium the free energy is, in general, a
non local functional thus implying the existence of correlations at
the macroscopic scale.  These correlations have been observed
experimentally \cite{DKS}
and appear to be a generic consequence of our
variational principle which can be reformulated as a time independent
Hamilton-Jacobi equation for the free energy. This is a functional
derivative equation whose independent arguments are the local
thermodynamic variables and which requires as input the transport
coefficients. 
The presence of long range correlations for out of equilibrium states
had been derived within kinetic theory, see \cite{DKS} and reference
therein, as well as within fluctuating hydrodynamics, see
e.g.\ \cite{S1}.

We believe that the theory we propose is a substantial improvement
with respect to the theory developed long ago by Onsager \cite{ONS1}
and then by Onsager-Machlup \cite{OMA} which applies to states close
to equilibrium, namely for linear evolution equations, and does not
really include the effect of nontrivial boundary reservoirs.
In principle, the theory we suggest should be applicable to real
systems, i.e.\ with nonlinear evolution equations and arbitrary
boundary conditions, where the diffusion is the dominant dynamical
mechanism.  We emphasize however that we assume a linear response with
respect to the external applied field. Of course, the Onsager theory
is recovered as first order approximation.

The basic principle of the theory here proposed is a variational
principle for the nonequilibrium thermodynamic functionals.
This principle has the following content.
Take the stationary state as the reference state and consider a
trajectory leading the system to a new state.
This trajectory can be realized by {imposing} a suitable
additional external field (in addition to the one already acting on
the system).
Then compute the work done by this extra field and
minimize it over all possible trajectories leading to the new state.
This minimal work is identified with the variation of the free energy
between the reference and the final state.
As well known in thermodynamics,
in the case of equilibrium states this definition agrees with the
standard one.

Our treatment is based on an approach developed by the authors in the
analysis of fluctuations in stochastic lattice gases
\cite{BDGJL1,BDGJL2,BDGJL3,BDGJL5,BDGJL6,BDGJL9,BGL}.  
For recent overviews on nonequilibrium phenomena see also
\cite{Derridareview,HS,MKS} and \cite{sew} for a different approach
which includes a discussion on long range correlations.

\bigskip
\noindent
The outline of the paper is the following.

In Section \ref{sec:2} we introduce the basic principles for a
self-contained macroscopic description of driven diffusive systems out
of equilibrium.
These systems are described by local thermodynamic variables
which evolve according to (nonlinear) diffusion equations with
boundary conditions and/or external fields which force the system out
of equilibrium. We emphasize that such evolution equations have the
structure of conservation laws.  The possibility of such description
is of course based on the assumption of local equilibrium.  We then
formulate mathematically the above mentioned variational principle for
the nonequilibrium free energy and characterize the optimal trajectory.
As we already remarked, our approach gives the usual equilibrium free
energy when we consider an equilibrium state.  In this connection,
we remark that the optimal trajectory provides a dynamical alternative
to the usual prescription of equilibrium thermodynamics to
calculate the free energy variation via a quasi static transformation.

In Section~\ref{sec:3} we discuss the notion of \emph{equilibrium}
from the standpoint of \emph{nonequilibrium} under the assumptions 
formulated is Section~\ref{sec:2}.
In particular we introduce the notion of macroscopic reversibility and
discuss its relationship, actually equivalence, to thermodynamic
equilibrium. An important outcome of our analysis is that the correct
definition of equilibrium, for the driven diffusive systems here
considered, is the vanishing of the currents.  Nonetheless, in
presence of external (electric) fields and boundary reservoirs an
equilibrium state can be highly inhomogeneous.  An example of such a
situation is provided by sedimentation equilibrium in gravitational
and centrifugal fields.  In spite of this, the free energy is a local
function of the state variables and coincides locally with the
equilibrium free energy in absence of external fields and boundary
driving. In particular there are no macroscopic correlations.

In Section~\ref{sec:4} we derive from the Hamilton-Jacobi equation the
generic existence of long range correlations in nonequilibrium states
and we discuss the conditions on the transport coefficients for their
appearance in the two point correlation function. We also establish a
simple criterion to determine whether density fluctuations are
positively or negatively correlated. 
Finally, from the Hamilton-Jacobi equation, 
by expanding around the stationary state, we derive a recursive equation
for the correlation functions of any order.

In Appendix~\ref{s:abc} we discuss a specific example, the \emph{ABC
  model} \cite{CDE,EKKM2} which does not satisfy all the assumptions
formulated in Section~\ref{sec:2}. In fact the ABC model has the
peculiarity to be a reversible system with a nonlocal free energy.
We show nonetheless that this free energy can be easily derived from
the Hamilton-Jacobi equation. 
This model has been brought to our attention by
D.\ Mukamel and J.L.\ Lebowitz. 


\section{Macroscopic systems out of equilibrium}
\label{sec:2}

We here introduce the thermodynamic description for out of
equilibrium driven diffusive systems which are characterized by
conservation laws. For simplicity of notation, we restrict to the case
of a single conservation law, e.g.\ the conservation of the mass. 
The system is in contact with boundary reservoirs,
characterized by their chemical potential $\lambda_0$, 
and under the action of an external field $E$.  
We denote by $\Lambda \subset \bb R^d$ the
region occupied by the system, by $x$ the macroscopic space
coordinates and by $t$ the time.  We next state our basic axioms and
their main implications.

\begin{itemize}
\item[1.]
\emph{The macroscopic state is completely described by
the local density $\rho=\rho(t,x)$ and the associated current $j=j(t,x)$.}

\item[2.]\emph{The macroscopic evolution is given by the continuity equation
\begin{equation}
\label{2.1}
\partial_t \rho + \nabla\cdot j = 0
\end{equation}
together with the constitutive equation
\begin{equation}
\label{2.2}
j = J(\rho) = - D(\rho) \nabla\rho + \chi(\rho) E
\end{equation}
where the \emph{diffusion coefficient} $D(\rho)$ and the \emph{mobility}
$\chi(\rho)$ are $d\times d$ positive matrices.
The transport coefficients $D$ and $\chi$ satisfy the local Einstein
relation
\begin{equation}
\label{ein_rel}
D(\rho) = \chi(\rho) \, f_0''(\rho)
\end{equation}
where $f_0$ is the equilibrium free energy of the homogeneous system.
The equations \eqref{2.1}--\eqref{2.2} have to be supplemented by the
appropriate boundary conditions on $\partial\Lambda$ due to the
interaction with the external reservoirs. Recalling that
$\lambda_0(x)$, $x\in\partial \Lambda$, is the chemical potential of
the external reservoirs, these boundary conditions are
\begin{equation}
\label{2.3}
f_0'\big(\rho(x) \big) = \lambda_0(x) \qquad\qquad x\in\partial
\Lambda
\end{equation}
We denote by $\bar\rho=\bar\rho(x)$, $x\in\Lambda$, the stationary
solution, assumed to be unique, of \eqref{2.1}, \eqref{2.2}, and \eqref{2.3}.}
\end{itemize}

In postulating the constitutive equation \eqref{2.2} we assumed a
linear response with respect to the external field $E$ and the
thermodynamic forces. On the other hand the transport coefficients
$D$ and $\chi$ are allowed to depend on the density $\rho$.
In the case the system has more than one component, say $n$, the
diffusion coefficient $D$ and the mobility become $nd\times nd$
matrices. Moreover, in view of Onsager reciprocity, the matrix $\chi$
is symmetric both in the space and in the component indices while $D$
is symmetric only in the space indices. In such a case the local Einstein
relation \eqref{ein_rel} is $D =  \chi \, R$ where
$R_{ij} = \partial_{\rho_i} \partial_{\rho_j} f_0$
does not depend on the space indices.
In the context of stochastic lattice gases, \eqref{2.1} and
\eqref{2.2} describe the evolution of the empirical density in the
diffusive scaling limit, see e.g.\ \cite{BDGJL9,KL,S};
the validity of the local Einstein relationship \eqref{ein_rel} can be
deduced from the local detailed balance of the underlying microscopic
dynamics, see e.g.\ \cite{S}. 
However, as we will see later, \eqref{ein_rel} it is also a
consequence of the locality of the free energy and our third postulate
once equilibrium is characterized within our theory.

To state the third postulate, we need some preliminaries.
Consider a time dependent variation $F= F(t,x)$ of the external field
so that the total applied field is $E+F$. The local current then becomes
$j=J^F(\rho) = J(\rho) + \chi(\rho) F$.
Given a time interval $[T_1,T_2]$, we compute the energy necessary to
create the extra current $J^F- J$ and drive the system along the
corresponding trajectory:
\begin{equation}
\label{2.4}
L_{[{T_1},{T_2}]}(F) =  
\int_{{T_1}}^{{T_2}} \! dt 
   \: \big\langle  \big[J^F(\rho^F) -
  J (\rho^F)\big] \cdot F  \big\rangle
= 
\int_{{T_1}}^{{T_2}} \! dt \: 
  \big\langle F  \cdot \chi(\rho^F) F \big\rangle
\end{equation}
where $\cdot$ is the scalar product in $\bb R^d$, 
$\langle \cdot \rangle$ is the integration over $\Lambda$,
and $\rho^F$ is the solution of the continuity equation with current
$j=J^F(\rho)$.


We define a \emph{cost functional} on the set of space
time trajectories as follows. Given a trajectory
$\hat\rho=\hat\rho(t,x),\,{t\in[T_1,T_2],x\in\Lambda}$, we set
\begin{equation}
\label{2.5}
I_{[{T_1},{T_2}]}(\hat\rho)
= \frac 14 \inf_{F\,: \: \rho^F= \hat\rho} \; L_{{[T_1,T_2]}}(F)
\end{equation}
namely we minimize over the {variations $F$} of the applied field
which produce the trajectory $\hat\rho$.  The introduction of this
functional in the context of fluid dynamics equations appears to be
new: it is not the total energy provided by the driving field
$F$ but only the energy necessary to create the extra current.  
In fact, $I$ (with the factor $1/4$) has a precise
statistical interpretation within the context of stochastic lattice
gases: it gives the asymptotics, as the number of degrees of freedom
diverges, of the probability of observing a space time fluctuation of
the empirical density \cite{BDGJL1,BDGJL2,BDGJL6,KL,KOV}. Note indeed
that if $\hat\rho$ solves the hydrodynamic equation \eqref{2.1},
\eqref{2.2}, \eqref{2.3} its cost vanishes.  We shall explain below
the choice of the factor $1/4$ in \eqref{2.5}.

We next show that the optimal $F$ in \eqref{2.5} is 
given by $F=\nabla\Pi$ where $\Pi : [T_1,T_2]\times\Lambda \to \bb R$
is the unique solution to the Poisson equation 
\begin{equation*}
- \nabla \cdot \big[ \chi(\hat\rho) \nabla \Pi \big] 
= \partial_t \hat\rho  + \nabla \cdot J(\hat\rho)  
\end{equation*}
which vanishes at the boundary of $\Lambda$ for any $t\in[T_1,T_2]$.
Indeed, by writing $F= \nabla \Pi + \tilde F$ for each $t\in [T_1,T_2]$ 
we have the orthogonal decomposition  
\begin{equation*}
\big\langle F  \cdot \chi(\hat\rho)  F \big\rangle = 
\big\langle \nabla \Pi \cdot \chi(\hat\rho)  \nabla \Pi
\big\rangle  
+ \big\langle \tilde F \cdot \chi(\hat\rho) \tilde F \big\rangle 
\end{equation*}
where we used that $\partial_t \hat\rho + \nabla\cdot J^F(\hat\rho)=0$. 
The above equation clearly shows that $\tilde F=0$ is the optimal choice. 
Hence
\begin{equation}
\label{I=}
I_{{[T_1,T_2]}} (\hat\rho) =
\frac 14 \int_{{T_1}}^{{T_2}}\!dt
\: \Big\langle
\big[ \partial_t \hat\rho  + \nabla \cdot J(\hat\rho)
\big] \, K(\hat\rho)^{-1}
\big[ \partial_t \hat\rho  + \nabla \cdot J(\hat\rho) \big] \Big\rangle
\end{equation}
where the positive operator $K(\hat\rho)$ is defined on functions $u
:\Lambda\to \bb R$ vanishing at the boundary $\partial \Lambda$ by
$K(\hat\rho) u = - \nabla \cdot\big( \chi(\hat\rho) \nabla u \big)$.

Our third postulate then characterizes the free energy $\mc F(\rho)$
of the system with a density profile $\rho=\rho(x),\,x\in\Lambda$, as
the minimal cost to reach, starting from the stationary profile
$\bar\rho$, the density profile $\rho$, in an infinitely long time
interval.

\begin{itemize}
\item[3.]\emph{The nonequilibrium free energy of the system is
\begin{equation}
\label{2.6}
\mc F (\rho) =
   \inf_{ \substack{ \hat\rho \, : \: \hat\rho({-\infty}) = \bar\rho \\
       \phantom{ \hat\rho \, : \:} \hat\rho({0}) =\rho} }
   I_{{[-\infty,0]}}(\hat\rho)
\end{equation}
}
\end{itemize}

As we shall see later, this is in fact proportional to the total work
done by the optimal external field $F$ along the optimal time
evolution to reach the density profile $\rho$.
The functional $\mc F$ has a precise statistical interpretation
within the context of (non equilibrium) stationary states of 
stochastic lattice gases: it gives the
asymptotics, as the number of degrees of freedom diverges, of the
probability of observing a static fluctuation of the empirical density
\cite{BDGJL1,BDGJL2,BDGJL6}. In this sense the functional $\mc F$
is the correct extension of the equilibrium free energy to 
non equilibrium thermodynamics.

As shown in \cite{BDGJL1,BDGJL2} the functional $\mc F$ is the maximal
solution of the infinite dimensional Hamilton-Jacobi equation
\begin{equation}
\label{HJeq}
\Big\langle  \nabla \frac{\delta\mc F}{\delta\rho} \cdot \chi(\rho)
\nabla \frac{\delta\mc F}{\delta\rho} \Big\rangle -
\Big\langle  \frac{\delta\mc F}{\delta\rho}
\: \nabla \cdot J(\rho) \Big\rangle  = 0
\end{equation}
where, for $\rho$ that satisfies \eqref{2.3}, $\delta\mc F /
\delta\rho$ vanishes at the boundary of $\Lambda$.  At the macroscopic
level this condition reflects the fact that we consider variations of
the density that do not change the boundary values. In the context of
stochastic lattice gases, the boundary condition \eqref{2.3} is
realized by superimposing to the bulk dynamics suitable birth and
death processes at the boundary; the probability of deviations of the
empirical density at the boundary is then shown to be
super-exponentially small in the scaling limit \cite{BDGJL2}. Note
however that in the stationary state the density is not fixed at the
boundary, i.e.\ $\mc F(\rho) <\infty$ even if the profile
$\rho=\rho(x)$ does not satisfy the boundary condition \eqref{2.3}.

The arbitrary additive constant on the maximal solution of
\eqref{HJeq} is determined by the condition $\mc F(\bar\rho)=0$. By
maximal solution we mean that any solution to \eqref{HJeq} (satisfying
$F(\bar\rho)=0$) is a lower bound for $\mc F$, as defined in
\eqref{2.6}. This bound is sharp.  Indeed, by considering the
functional in \eqref{I=} as an action functional in variables $\hat
\rho$ and $\partial_t \hat\rho$ and performing a Legendre transform,
the associated Hamiltonian is
\begin{equation}
\label{Ham}
\mc H (\rho,\Pi) = \Big\langle  \nabla \Pi \cdot \chi(\rho)
\nabla  \Pi \Big\rangle +  \Big\langle  \nabla\Pi  \cdot J(\rho) \Big\rangle
\end{equation}
where the \emph{momentum} $\Pi$ vanishes at the boundary of
$\Lambda$.  By noticing that the stationary solution of the
hydrodynamic equation corresponds to the equilibrium point
$(\bar\rho,0)$ of the system with Hamiltonian $\mc H$ and $\mc H
(\bar\rho,0)=0$, the equation \eqref{HJeq} is the Hamilton-Jacobi
equation $\mc H \big(\rho,\delta \mc F/ \delta \rho\big) =0$.

The optimal trajectory $\rho^*$ for the variational problem
\eqref{2.6} is characterized as follows.  Let
\begin{equation}
\label{2.7}
J^*(\rho) = - 2\chi(\rho) \nabla \frac{\delta\mc F}{\delta\rho}  - J(\rho)
\end{equation}
then $\rho^*$ is the time reversal of the solution to
\begin{equation}
\label{2.7b}
\partial_t \rho + \nabla \cdot J^*(\rho) =
\partial_t \rho + \nabla \cdot \Big\{
D(\rho) \nabla \rho -
\chi(\rho) \Big[ E+  2 \, \nabla \frac{\delta\mc F}{\delta\rho} \Big] \Big\}
=  0
\end{equation}
with the boundary condition \eqref{2.3}. We refer to
\cite{BDGJL1,BDGJL2,BDGJL6} for the microscopic interpretation of $J^*$
in terms of the time reversed dynamics.
In particular, the optimal external field $F$ forcing the system
on the time evolution $\rho^*(t)$ is so that $J^F(\rho)=-J^*(\rho)$,
namely
\begin{equation}\label{alb0}
F=2 \, \nabla \frac{\delta \mc F}{\delta \rho}
\end{equation}
This characterization of $\rho^*$ is obtained
as follows.
Let $\mc F$ be the maximal solution of the Hamilton-Jacobi equation and
$J^*$ as in \eqref{2.7}.
Fix a time interval ${[T_1,T_2]}$ and a
path $\hat \rho (t)$, $t\in {[T_1,T_2]}$. We claim that
\begin{eqnarray}
\label{I=I*}
\nonumber
&& I_{{[T_1,T_2]}} (\hat\rho) = \mc F\big(\hat\rho({T_2})\big) 
-\mc F\big(\hat\rho({T_1}) \big)
\\
&& \qquad\qquad
+ \frac 14 \int_{{T_1}}^{{T_2}}\!dt\:
\: \Big\langle
\big[ \partial_t \hat\rho  - \nabla \cdot J^*(\hat\rho)
\big] \, K(\hat\rho)^{-1}
\big[ \partial_t \hat\rho  - \nabla \cdot J^*(\hat\rho) \big]
\Big\rangle \qquad
\end{eqnarray}
as can be shown by a direct computation using \eqref{I=},
the Hamilton-Jacobi equation \eqref{HJeq} and the definition
\eqref{2.7} of $J^*$. 
From the identity \eqref{I=I*} we immediately deduce
that the optimal path for the variational problem
\eqref{2.6} is the time reversal of the solution to \eqref{2.7b}.

\medskip
The classical thermodynamic setting considers only spatially
homogeneous system in which $\rho$ does not depend on $x$.  In this
case the variation of the free energy between $\bar\rho$ and $\rho$ is
the minimal work required to drive the system from $\bar\rho$ to
$\rho$, which is realized by a \emph{quasi static} transformation, see
e.g.\ \cite{LL}.  For an equilibrium inhomogeneous density profile
$\rho$, by a quasi static transformation we understand the following:
pick a macroscopic point $x$ and consider a macroscopically small
volume around $x$, perform quasi static transformation from
$\bar\rho(x)$ to $\rho(x)$, repeat independently for all $x$ and sum
the results.  Of course, if the system is in contact with a reservoir
also the contribution describing the interaction with the reservoir
has to be included.

We next show that according to definition \eqref{2.6}, the free energy
$\mc F(\rho)$ is proportional to the work done by the optimal external
field $F$ on the system along the optimal trajectory $\rho^*$.  We
emphasize {that} this is not a quasi static transformation in the
above sense, but it is the solution of the hydrodynamic equation
perturbed by the optimal external field.  As we shall discuss later,
the definition \eqref{2.6} of the free energy (with the factor $1/4$
in \eqref{2.5}) agrees with the thermodynamic one for equilibrium
states, but in general, due to presence of long range correlations,
the variation obtained along a quasi static transformation as defined
above gives a different result.

Indeed, since $\rho^*$ is the time reversal of the solution to
\eqref{2.7b}, using \eqref{alb0} we deduce
\begin{eqnarray*}
\mc F(\rho)-\mc F(\bar\rho) 
&= &\int_{-\infty}^0 \!dt \, \Big\langle \frac{\delta\mc F}{\delta \rho},
\partial_t\rho^* \Big\rangle
\\
&=& \int_{-\infty}^0 \!dt\, \Big\langle \frac{\delta\mc F}{\delta \rho},
\nabla \cdot J^*(\rho^*) \Big\rangle
=\frac 12 \int_{-\infty}^0 \!dt \, \Big\langle (-J^*(\rho^*))\cdot F \Big\rangle
\end{eqnarray*}
Namely, the variation $\mc F(\rho)-\mc F(\bar\rho)$ is proportional the
work done by the external field $F$ along the optimal trajectory
$\rho^*$. The factor $1/2$ above is connected to our definition of the
transport coefficients and the associated Einstein relation
\eqref{ein_rel}.  

\medskip
As shown in the next section, equilibrium states are characterized by
$J=J^*$. In this case the optimal path for the variational problem
\eqref{2.6} is the time reversal of the solution to the original
hydrodynamic equation \eqref{2.1}--\eqref{2.2}.
In \cite{BDGJL2} we called such a symmetry property the
Onsager-Machlup time reversal symmetry.

\section{Characterizations of equilibrium states}
\label{sec:3}

We define the system to be in
\emph{equilibrium} if and only if the current in the stationary profile
$\bar\rho$ vanishes, i.e.\ $J(\bar\rho) = 0$.
A particular case is that of a \emph{homogeneous equilibrium state},
obtained by setting the external field $E=0$ and choosing a constant
chemical potential potential at the boundary, i.e.\ $\lambda_0(x) =
\bar\lambda$. Let $\bar\rho=\mathrm{const.}$ be the
equilibrium density, i.e.\ $\bar\rho$
solves $\bar\lambda= f_0'(\bar\rho)$.
It is then readily seen that the functional $\mc F$ defined
in \eqref{2.6} is given by
\begin{equation}
\label{3.14}
\mc F(\rho) = \int_{\Lambda} \!dx \:
\big\{
f_0\big(\rho(x)\big) - f_0(\bar\rho) -
{\bar\lambda}
\big[ \rho(x) -\bar\rho\big]
\big\}
\end{equation}
in which the first difference is the variation of the free energy
$f_0$ while the second term is due to the interaction with the
reservoirs. The variational definition \eqref{2.6} of
the functional $\mc F$ gives, when applied to equilibrium states, the
variation of the equilibrium free energy \eqref{3.14}. This shows that the 
choice of the factor $1/4$ in \eqref{2.5} is appropriate.

We next show that also for a non homogeneous equilibrium, characterized
by a non constant stationary profile $\bar\rho(x)$ such that
$J(\bar\rho)=0$ the free energy functional $\mc F$ can be explicitly
computed.
Let
\begin{equation*}
f(\rho,x) =
\int_{\bar\rho(x)}^\rho \!dr \int_{\bar\rho(x)}^r \!dr' \: f_0''(r')
= f_0(\rho) - f_0(\bar\rho(x)) - f_0'\big(\bar\rho(x)\big) \big[ \rho
-\bar\rho(x)\big] 
\end{equation*}
we claim that 
the maximal solution of the Hamilton-Jacobi equation is 
\begin{equation}
\label{Floc}
\mc F(\rho) = \int_\Lambda \!dx \: f\big( \rho(x), x\big)
\end{equation}
Indeed
\begin{equation}
\label{varf}
\frac{\delta\mc F}{\delta\rho(x)} = f_0'(\rho(x))-f_0'(\bar\rho(x))
\end{equation}
so that, by an integration by parts,
\begin{eqnarray*}
&&
\Big\langle \nabla\big[ f_0'(\rho) - f_0'(\bar\rho) \big] \cdot
\chi(\rho) \nabla\big[ f_0'(\rho) - f_0'(\bar\rho) \big] \Big\rangle
\\
&& \qquad\qquad
+ \Big\langle \big[ f_0'(\rho) - f_0'(\bar\rho) \big]
\nabla\cdot \Big[D(\rho)\nabla\rho - \chi(\rho)E \Big] \Big\rangle
\\
&&\qquad = \Big\langle \nabla\big[ f_0'(\rho) - f_0'(\bar\rho) \big] \cdot
\chi(\rho) \Big[ \nabla f_0'(\bar\rho) - E \Big] \Big\rangle
=0
\end{eqnarray*}
where we used \eqref{ein_rel} and  $\nabla f_0'(\bar\rho) - E =
-\chi(\bar\rho)^{-1} J(\bar\rho)= 0$.
Therefore the functional $\mc F$ in \eqref{Floc} satisfies the
Hamilton-Jacobi equation \eqref{HJeq}. 
To show it is the maximal solution,
recalling \eqref{I=}, simple computations show that
\begin{eqnarray}
\nonumber
\label{rev}
I_{[T_1,T_2]} (\hat \rho) &=&
\mc F\big(\hat\rho(T_2)\big) - \mc F\big(\hat\rho(T_1)\big)
\\
&&+ \frac 14 \int_{T_1}^{T_2}\!dt \: \Big\langle
\big[ \partial_t \hat\rho - \nabla \cdot J(\hat\rho)\big] \, K(\hat\rho)^{-1}
\big[ \partial_t \hat\rho  - \nabla \cdot J(\hat\rho)\big]
\Big\rangle \quad\quad
\end{eqnarray}
which clearly implies the maximality of $\mc F$.

\medskip
We emphasize that the above argument depends crucially on the 
structure \eqref{2.2} of the current and on validity
of the local Einstein relation \eqref{ein_rel}. 
In Appendix~\ref{s:abc} we discuss the example of the ABC model
\cite{CDE,EKKM2} in which these conditions are not met
and the free energy is not a local functional even if $J(\bar\rho)=0$.
Another example is given by the anisotropic zero-range process
\cite{joeletal,vB} for which - as discussed in \cite{vB} - the
local Einstein relation is violated.

\medskip
We next show that the condition $J(\bar\rho)=0$
is equivalent to either one of the following statements.

\begin{itemize}
\item[--]{
There exists a function $\lambda: \Lambda \to \bb R$
such that
\begin{equation}
\label{cond_grad}
E(x)= \nabla\lambda (x)\,, \quad x\in \Lambda
\qquad \quad
\lambda(x) = \lambda_0 (x) \,, \quad x\in \partial\Lambda
\end{equation}
}
\item[--]{The system is \emph{macroscopically reversible} in the sense
  that for each profile $\rho$ we have $J^*(\rho) = J(\rho)$.}
\end{itemize}

We emphasize that the notion of macroscopic reversibility does not
imply that an underlying microscopic model satisfies the detailed
balance condition. Indeed, as it has been shown by explicit examples
\cite{GJL1,GJL2}, there are non reversible microscopic models which are
macroscopically reversible.

We start by showing that $J(\bar\rho) = 0$ if and only if
\eqref{cond_grad} holds.  From the local Einstein relation \eqref{ein_rel}
and $J(\bar\rho) = 0$ we deduce
\begin{equation*}
E(x)= f''_0\big(\bar{\rho}(x) \big) \nabla\bar{\rho}(x)
= \nabla f'_0(\bar{\rho}(x))
\end{equation*}
hence \eqref{cond_grad}.
Conversely, let the external field $E$ be such
that \eqref{cond_grad} holds.  Since $f_0''$ is positive the function
$f_0'$ is invertible and we can define
$\bar{\rho}(x)=(f_0')^{-1}\big(\lambda(x) \big)$. The profile $\bar\rho$
satisfies \eqref{2.3} as well as $J(\bar{\rho})=0$, consequently also
$\nabla \cdot J(\bar{\rho})=0$ so that it is the unique stationary
profile.

We next show that $J(\bar\rho)=0$ if and only if $J(\rho)=J^*(\rho)$.
By evaluating the Hamilton-Jacobi equation for $\rho=\bar\rho$ we
deduce $ \nabla \big[\delta \mc F (\bar\rho) / \delta \rho \big]
=0$. From \eqref{2.7} we then get $J(\bar\rho)=0$.
To show the converse implication, note that if $J(\bar\rho)=0$
then $\mc F$ is given by \eqref{Floc}. We deduce
\begin{equation*}
\chi(\rho) \nabla \frac {\delta \mc F}{\delta \rho} =
D(\rho)\nabla\rho -  \chi(\rho) E = - J(\rho)
\end{equation*}
hence, recalling \eqref{2.2} and \eqref{2.7}, $J(\rho)=J^*(\rho)$.

\medskip
We also note that macroscopic reversibility $J(\rho)=J^*(\rho)$
implies the invariance of the Hamiltonian $\mc H$ in \eqref{Ham}
under the time reversal symmetry, see \cite{BDGJL2},
$(\rho,\Pi)\mapsto \big(\rho, \delta \mc F/\delta \rho -\Pi \big)$,
where $\mc F$ is the maximal solution of the Hamilton-Jacobi equation
\eqref{HJeq}.
As remarked before, macroscopic reversibility implies the
Onsager-Machlup time reversal symmetry,
on the other hand we shall give 
at the end of this section
an example of a nonequilibrium system for which such a symmetry
holds. Hence macroscopic reversibility is a stronger condition.

So far we have assumed the local Einstein relation and we have shown that
-for equilibrium systems- it implies \eqref{Floc}. Conversely,
we now show that macroscopic reversibility and \eqref{Floc}
imply the local Einstein relation \eqref{ein_rel}.
By writing explicitly $J(\rho)=J^*(\rho)$ we obtain
\begin{equation}
\label{einbis}
-\big[ \chi(\rho) R(\rho)-D(\rho) \big] \nabla \rho
+ \chi(\rho) \big[ R(\bar \rho)- \chi^{-1}(\bar \rho)D(\bar \rho)
\big]\nabla \bar \rho = 0
\end{equation}
where $R$ is the second derivative of $f_0$ in the case of
one-component systems while
$R_{ij}=\partial_{\rho_i} \partial_{ \rho_j} f_0$  for multi-component systems.
In \eqref{einbis} we used, besides \eqref{Floc}, $J(\bar \rho)=0$ to eliminate
$E$. Note that $J(\bar \rho)=0$  follows from the Hamilton-Jacobi
equation and  $J(\rho)=J^*(\rho) $ without further assumptions.
Since $\rho$ and $\nabla \rho$ are arbitrary the local Einstein relation
$D = \chi R$ follows from \eqref{einbis}.

\medskip
We have defined the macroscopic reversibility as the identity between
the currents $J(\rho)$ and $J^*(\rho)$. We emphasize that this is not
equivalent to the identity between $\nabla \cdot J(\rho)$ and $\nabla
\cdot J^*(\rho)$.  Indeed, we next {give an example of}
a non reversible system, i.e.\ {with} $J(\bar\rho) \neq 0$, such that the optimal
trajectory for the variational problem \eqref{2.6} is the time
reversal of the solution to the hydrodynamic equation
\eqref{2.1},\eqref{2.2}, and \eqref{2.3}.

Let $\Lambda = [0,1]$, $D(\rho)=\chi(\rho)=1$,
$\lambda_0(0)=\lambda_0(1)=\bar\lambda$, and a constant external field
$E\neq 0$. In this case hydrodynamic evolution of the density is given
by the heat equation independently of the field $E$.
The stationary profile is $\bar\rho = \bar\lambda$, the associated
current is  $J(\bar\rho) = E \neq 0$.  By a computation analogous
to the one leading to {\eqref{rev}}, we easily get that
\begin{equation*}
\mc F(\rho) = \frac 12 \int_0^1\!dx \: \big[ \rho(x) - \bar\rho \big]^2
\end{equation*}
and the optimal trajectory for the variational problem \eqref{2.6} is
the time reversal of the solution to the heat equation.
On the other hand $J(\rho)= - \nabla \rho +E$ while
$J^*(\rho)= -\nabla \rho -E$

\medskip
We remark that, even if the free energy $\mc F$ is a non local
functional, the equality $J(\rho)=J^*(\rho)$ implies that the
thermodynamic force $ \nabla \delta \mc F / \delta \rho$ is local.  
Moreover, the Hamilton-Jacobi equation reduces to the statement
\begin{equation}
\label{ein?}
J(\rho) =  - \chi(\rho) \nabla \frac{\delta \mc F}{\delta \rho}   (\rho)
\end{equation}
This identity, viewed as a special case of \eqref{2.7}, represents the
general form -- for reversible systems -- of the relationship between
currents and thermodynamic forces.  It holds both when the free energy
is local and nonlocal. For reversible systems it is therefore enough
to integrate \eqref{ein?} to compute the free energy.  An example of
such a situation with a nonlocal free energy, illustrated in
Appendix~\ref{s:abc}, is provided by the ABC model on a ring with
equal densities \cite{CDE,EKKM2}.

\section{Correlation functions}
\label{sec:4}

In classical equilibrium thermodynamics the second derivatives of the
the thermodynamic potentials like the free energy are  response
coefficients like the compressibility. In a statistical mechanics
framework they are identified with the Fourier transform evaluated at
zero of correlation functions of the underlying Gibbs measure.
Analogous interpretation is given to the higher order derivatives.

As we defined the (nonequilibrium) free energy as a functional on the
set of density profiles, we can obtain the (nonequilibrium) density
correlations functions of arbitrary order in terms of the functional
derivatives of $\mc F$. In general the functional $\mc F$ cannot be
written in a closed form, but - by a suitable perturbation theory on the
Hamilton-Jacobi equation \eqref{HJeq} - we can derive such
correlations functions.
In this section we first discuss the two-point correlation and establish a
criterion to decide whether the density fluctuations are positively or
negatively correlated. Then, we deduce a recursive equation for the
correlation function of any order.

We remark that we are concerned only with \emph{macroscopic
  correlations} which are a generic feature of nonequilibrium models.
Microscopic correlations which decay as a summable power law 
disappear at the macroscopic level.

We introduce the \emph{pressure} functional as the Legendre
transform of free energy $\mc F$
\begin{equation*}
\mc G(h) = \sup_\rho\big\{\langle h \rho\rangle -\mc F(\rho)\big\}
\end{equation*}
By Legendre duality
{
we have the change of variable formulae
$h=\frac{\delta\mc F}{\delta\rho},\,\rho=\frac{\delta\mc G}{\delta h}$,
so that
}
the Hamilton-Jacobi equation \eqref{HJeq}
can then be rewritten in terms of $\mc G$ as
\begin{equation}
\label{s2}
\Big\langle \nabla h \cdot
\chi\Big(\frac{\delta\mc G}{\delta h} \Big)\nabla h
\Big\rangle
- \Big\langle \nabla h \cdot
 D \Big(\frac{\delta\mc G}{\delta h}\Big) \nabla\frac{\delta\mc
  G}{\delta h}  {-} \chi\Big(\frac{\delta\mc G}{\delta h}\Big)E
\Big\rangle = 0
\end{equation}
where $h$ vanishes at the boundary of $\Lambda$.
As for equilibrium systems, $\mc G$ is the generating functional of the correlation
functions. In particular by defining
\begin{equation*}
C(x,y) = \frac{\delta^2\mc G(h)}{\delta h(x) \delta h(y)} \,{\Big|_{h=0}}
\end{equation*}
we have, since $\mc F$ has a minimum at $\bar\rho$,
\begin{equation*}
\mc G(h) = \langle h,\bar\rho\rangle + \frac12 \langle
h,Ch\rangle + o(h^2)
\end{equation*}
or equivalently
\begin{equation*}
\mc F(\rho) = \frac12
\langle(\rho-\bar\rho),C^{-1}(\rho-\bar\rho)\rangle +
o((\rho-\bar\rho)^2)
\end{equation*}

By expanding the Hamilton-Jacobi equation \eqref{s2}
to the second order in $h$, and using that
$\delta \mc G /\delta h(x) = \bar\rho (x) + C h(x) +o(h^2)$,
we get the following equation for $C$
\begin{equation}
\label{s4}
\Big\langle \nabla h \cdot \Big[
\chi(\bar\rho)\nabla h -
\nabla(D(\bar\rho)Ch)+\chi^\prime(\bar\rho)(Ch)E \Big]\Big\rangle = 0
\end{equation}

We now make the change of variable
$$
C(x,y)=C_{\mathrm{eq}}(x)\delta(x-y)+B(x,y)
$$
where $C_{\mathrm{eq}}(x)$ is the equilibrium covariance. By using
\eqref{ein_rel} we deduce that
$$
C_{\mathrm{eq}}(x)= D^{-1}(\bar\rho(x))\chi(\bar\rho(x))
$$
Equation \eqref{s4} for the correlation function then gives the
following equation for $B$
\begin{equation}\label{s5}
\mc L^\dagger B(x,y) = \alpha(x)\delta(x-y)
\end{equation}
where $\mc L^\dagger$ is the formal adjoint of the
elliptic operator $\mc L=L_x+L_y$ given by, using the usual convention
that repeated indices are summed,
\begin{equation}\label{lx}
L_x = D_{ij}(\bar\rho(x))\partial_{x_i}\partial_{x_j} +
\chi^\prime_{ij}(\bar\rho(x)) E_j(x)\partial_{x_i}
\end{equation}
and
\begin{equation*}
\alpha(x) = \partial_{x_i}\big[
\chi^\prime_{ij}\big(\bar\rho(x)\big)\,
D^{-1}_{jk}\big(\bar\rho(x) \big)\bar J_{k}(x) \big]
\end{equation*}
where we recall
$\bar J= J (\bar\rho)=- D(\bar\rho(x))\nabla\bar\rho(x)
+\chi(\bar\rho(x))E(x)$
is the macroscopic current in the stationary profile.

In agreement with the discussion in Section~\ref{sec:3},
for equilibrium systems $\bar J=0$ so that we have
$\alpha=0$, hence $B=0$, namely there are no long range correlations
and $C(x,y)=C_{\mathrm{eq}}(x)\delta(x-y)$.
Moreover, since $\mc L$ is an elliptic operator 
(i.e. it has a
negative kernel), the sign of $B$ is determined by the sign of $\alpha$:
if $\alpha(x) \ge 0,\,\forall x$, then $B(x,y) \le 0,\, \forall x,y$,
while if $\alpha(x) \le 0,\,\forall x$, then $B(x,y) \ge 0,\, \forall
x,y$. 
For example, consider the following special case. The system is one-dimensional,
$d=1$, the diffusion coefficient is constant, i.e.\ $D(\rho)=D_0$,
the mobility $\chi(\rho)$ is a quadratic function of $\rho$, and there
is no external field, $E=0$. Then
\begin{equation}
\label{opp}
B(x,y) = - \frac 1{2D_0} \chi'' (\nabla \bar\rho)^2 \Delta^{-1}(x,y)
\end{equation}
where $\Delta^{-1}(x,y)$ is the Green function of the Dirichlet
Laplacian. Two remarkable models, the symmetric exclusion process,
where $\chi(\rho)=\rho(1-\rho)$, and the KMP process, where
$\chi(\rho)=\rho^2$, meet the above conditions. Then \eqref{opp} shows that 
their correlations have opposite signs. 

\bigskip
{
We next derive, using the Hamilton-Jacobi equation \eqref{s2},
a recursive formula for the $n$-point correlation function $C_n(x_1,\dots,x_n)$.
This is defined in terms of the pressure functional $\mc G$ as
\begin{equation}\label{alb5}
C_n(x_1,\dots,x_n)=
\frac{\delta^n\mc G}{\delta h(x_1)\cdots\delta h(x_n)}\,\Big|_{h=0}
\end{equation}
so that $C_1(x)=\bar\rho(x)$ and $C_2$ is the two-point correlation
function discussed above. 
Note that for equilibrium systems the free energy is given by
\eqref{Floc}, hence we have no long range correlations and the
$n$-point correlation function is
\begin{equation}\label{npeq}
C_{n,\mathrm{eq}}(x_1,\dots,x_n)=c_n(\bar\rho(x_1))\delta(x_1-x_2)
\cdots\delta(x_1-x_n)
\end{equation}
where $c_n(\bar\rho)=\frac{d^n}{dx^n}(f_0')^{-1}(x)\big|_{x=f_0'(\bar\rho)}$.

By expanding the functional derivative of $\mc G$ we get 
\begin{equation}\label{alb6}
\frac{\delta \mc G(h)}{\delta h(x_1)} = \bar\rho(x_1) 
+ \sum_{n\geq1} \frac1{n!} \mc G_n(h;x_1)
\end{equation}
where
\begin{equation}\label{alb7}
\mc G_n(h;x_1) = 
\int_\Lambda dx_2\dots dx_{n+1}
h(x_2)\dots h(x_{n+1}) C_{n+1}(x_1,x_2,\dots,x_{n+1})
\end{equation}
If we consider the terms of order $h^{n+1}$ in the Hamilton-Jacobi
equation \eqref{s2} we get, using \eqref{alb6} and after some
algebraic manipulations,
\begin{eqnarray*}
\Big\langle\nabla h\cdot\Bigg[
\sum_{\substack{i_1,i_2,\dots\geq0 \\ \sum_{k}k i_k=n-1}}
\frac{1}{i_1!(1!)^{i_1}i_2!(2!)^{i_2}\cdots} 
 \chi^{(\sum_k i_k)}(\bar\rho)
(\mc G_1)^{i_1}(\mc G_2)^{i_2}\dots\nabla h \nonumber\\
-\sum_{\substack{i_1,i_2,\dots\geq0 \\ \sum_{k}k i_k=n}}
\frac{1}{i_1!(1!)^{i_1}i_2!(2!)^{i_2}\cdots} 
\nabla\Big( D^{(\sum_k i_k-1)}(\bar\rho)
(\mc G_1)^{i_1}(\mc G_2)^{i_2}\dots\Big) \\
+\sum_{\substack{i_1,i_2,\dots\geq0 \\ \sum_{k}k i_k=n}}
\frac{1}{i_1!(1!)^{i_1}i_2!(2!)^{i_2}\cdots} 
\chi^{(\sum_k i_k)}(\bar\rho)
(\mc G_1)^{i_1}(\mc G_2)^{i_2}\dots E\Bigg]\Big\rangle = 0\nonumber
\end{eqnarray*}
where, for $f$ given by either $\chi$ or $D$, the notation
$f^{(k)}$ stands for the $k$-th derivative of $f$.  

Since the above identity holds for every function $h$ vanishing at the
boundary of $\Lambda$, we get from it a recursive formula for the
$n$-point correlation functions which generalizes \eqref{s5}.  We
start by introducing some notation.  For $n\geq1$, we let $\mc
L^\dagger_{n+1}$ be the formal adjoint of the operator $\mc
L_{n+1}=\sum_{k=1}^{n+1}L_{x_k}$, where $L_x$ is defined in
\eqref{lx}.  Given functions $f$ and $g$ in $n+1$ and $m+1$ variables
respectively, }{ their 1st shuffle product $f\#_1g$ is, by definition,
  the following function in $m+n+1$ variables
\begin{eqnarray*}
&&f\#_1g(x_1,\dots,x_{m+n+1})
\\
&&\qquad
=\;\frac{1}{(m+n)!}\sum_\pi f(x_1,x_{\pi_2},\dots,x_{\pi_{n+1}})
g(x_1,x_{\pi_{n+2}},\dots,x_{\pi_{m+n+1}})
\qquad
\end{eqnarray*}
where the sum is over all permutations of the indices $\{2,3,\dots,m+n+1\}$.
Clearly $\#_1$ is a commutative associative product, and $f\#_1g$ is symmetric
in the variables $x_2,\dots,x_{m+n+2}$.
Furthermore, given a function $f$ in $n$ variables, we denote by
$f^{sym}$ its symmetrization, namely
\begin{equation*}
f^{sym}(x_1,\dots,x_n)=\frac1{n!}\sum_{\pi\in S_n}f(x_{\pi_1},\dots,x_{\pi_{n}})
\end{equation*}
Given a sequence $\vec{\iota}=(i_1,i_2,i_3,\dots)$ of non negative integers with
only finitely many non zero entries, we also denote
\begin{eqnarray*}
\Sigma(\vec \iota) &=& i_1+i_2+\dots=\sum_ki_k\\
N(\vec \iota) &=& i_1+2i_2+\dots=\sum_kki_k\\
K(\vec \iota) &=& i_1!(1!)^{i_1}i_2!(2!)^{i_2}\dots=\prod_ki_k!(k!)^{i_k}
\end{eqnarray*}
Finally, for $\vec \iota$ with $N(\vec \iota)=n$, we let
$$
C_{\vec \iota}(x_1,\dots,x_{n+1})
= \big(C_2^{\#_1 i_1}\#_1C_3^{\#_1 i_2}\#_1\cdots\big)(x_1,x_2,\dots,x_{n+1})
$$
Then the recursive equation satisfied by the $n+1$-point correlation
function 
reads
\begin{eqnarray}
\label{alb8}
&& \frac1{(n+1)!}\mc L_{n+1}^\dagger C_{n+1}(x_1,x_2,\dots,x_{n+1}) \\
&&
=\Bigg\{
\sum_{\substack{\vec \iota \\ N(\vec \iota)=n-1}}
\frac{1}{K(\vec \iota)} 
\nabla_{x_1}\cdot
\Big(
\chi^{(\Sigma(\vec\iota))}(\bar\rho(x_1))
C_{\vec\iota}(x_1,\dots,x_n)\nabla_{x_1} \delta(x_1-x_{n+1})
\Big)
\nonumber\\
&& 
-\sum_{\substack{\vec \iota \\ N(\vec \iota)=n,i_n=0}}
\frac{1}{K(\vec \iota)} 
\nabla_{x_1}\cdot\nabla_{x_1}
\Big(
D^{(\Sigma(\vec\iota)-1)}(\bar\rho(x_1))
C_{\vec\iota}(x_1,\dots,x_{n+1}) 
\Big) 
\nonumber\\
&&
+\sum_{\substack{\vec \iota \\ N(\vec \iota)=n,i_n=0}}
\frac{1}{K(\vec \iota)} 
\nabla_{x_1}\cdot
\Big(
\chi^{(\Sigma(\vec\iota))}(\bar\rho(x_1))
C_{\vec\iota}(x_1,\dots,x_{n+1}) E(x_1)
\Big)
\Bigg\}^{sym}
\nonumber 
\end{eqnarray}
It is easy to check that, for $n=1$, equation \eqref{alb8} 
reduces to \eqref{s5}. }
Moreover, in the case of equilibrium states the solution to
\eqref{alb8} is \eqref{npeq}. We refer to \cite[\S~4.5]{BDGJL2} for
the application to the boundary driven simple exclusion process.

\appendix
\section{The ABC model}
\label{s:abc}

We here consider - both from a microscopic and macroscopic point of
view - a model with two conservation laws.  Given an integer $N\ge 1$
let $\bb Z_N = \{1,\dots,N\}$ be the discrete ring with $N$ sites so
that $N+1\equiv 1$.  The microscopic space state is given by
$\Omega_N= \{A,B,C\}^{\bb Z_N}$ so that at each site $x\in\bb Z_N$ the
occupation variable, denoted by $\eta_x$, take values in the set
$\{A,B,C\}$; one may think that $A,B$ stand for two different species
of particles and $C$ for an empty site.  Note that this state space
takes into account an exclusion condition: at each site there is at
most one species of particles.

We first consider a weakly asymmetric dynamics that fits in the
framework discussed in Section~\ref{sec:2} that is defined
by choosing the following transition rates. If the occupation variables
across the bond $\{x,x+1\}$ are $(\xi,\zeta)$, they are exchanged to
$(\zeta, \xi)$ with rate 
$c^E_{x,x+1} = \exp\{ (E_{\xi} - E_{\zeta})/ (2 N) \}$ for
fixed constant external fields $E_A,E_B,E_C$. This choice satisfies
the so-called \emph{local detailed balance} condition that, in general,
requires the weakly asymmetric rates $c_{x,x+1}^E$ to satisfy 
\begin{eqnarray}
\label{ldb}
&&\!\!\!\!\!\!\!\!\! c^E_{x,x+1} (\eta^{x,x+1}) = c^E_{x,x+1} (\eta) 
\\
&&\!\!\!\!
\nonumber \times
\exp\Big\{ \nabla_{x,x+1} \Big[ H (\eta) -  \frac 1N 
\sum_x   \big(E_A x \id_A(\eta_x) +E_B x \id_B(\eta_x) +E_C x \id_C(\eta_x) \big)
\Big]\Big\} 
\end{eqnarray}
where $\eta^{x,x+1}$ is the configuration obtained from $\eta$ by
exchanging the occupation variables in $x$ and $x+1$, 
$\nabla_{x,x+1} f(\eta) = f(\eta^{x,x+1}) -f(\eta)$, and $H(\eta)$ is
the energy of the configuration $\eta$, which is constant in
this model.

Following \cite{DPS,KL,KOV}, the hydrodynamic equations  for the densities of
$A$ and $B$ particles are given by
\begin{equation}
\label{hy2}
  \partial_t
\left(
\begin{array}{c}
\rho_A  \\
\rho_B
\end{array}
\right)
= \Delta
\left(
\begin{array}{c}
\rho_A  \\
\rho_B
\end{array}
\right)
- \nabla \cdot
\left(
  \begin{array}{cc}
    \rho_A (1-\rho_A)   & - \rho_A \rho_B \\
     - \rho_A \rho_B    &  \rho_B (1-\rho_B)
  \end{array}
\right)
\left(
\begin{array}{c}
E_A-E_C  \\
E_B-E_C
\end{array}
\right)
\end{equation}
of course the density of $C$ particles is then $\rho_C=1-\rho_A-\rho_B$.

The functional $I_{[T_1,T_2]}$ in \eqref{I=} with $D = \id$ and
mobility
\begin{equation}
\label{chiabc}
\chi (\rho_A,\rho_B) =
\left(
  \begin{array}{cc}
    \rho_A (1-\rho_A)   & - \rho_A \rho_B \\
     - \rho_A \rho_B    &  \rho_B (1-\rho_B)
  \end{array}
\right)
\end{equation}
is the dynamical large functional associated to this
model. The free energy is the maximal solution of the Hamilton-Jacobi
equation \eqref{HJeq} which can be easily computed. Namely,
\begin{eqnarray}
  \label{qps=}
&&
\nonumber
  \mc F_{m_A,m_B}^0(\rho_A,\rho_B) =\int\!dx \,\Big[
    \rho_A \log\frac{\rho_A}{m_A}+\rho_B \log\frac{\rho_B}{m_B}
\\
&&\phantom{  \mc F_{m_A,m_B}(\rho_A,\rho_B) =\int\!dx \,\Big[ \;}
  +(1-\rho_A-\rho_B) \log\frac{1-\rho_A-\rho_B}{1-m_A-m_B} \Big]
\qquad\qquad
\end{eqnarray}
where $\int\!dx\, \rho_A = m_A$ and $\int\!dx\, \rho_B = m_B$.
If $E_A$, $E_B$ and $E_C$ are not all equal, this model is a
nonequilibrium model nevertheless, in view of the periodic boundary
conditions, its free energy is independent of the external
field; see \cite[\S~3.4]{BDGJL9}.

\medskip
We next discuss a different choice of the weakly asymmetric
perturbation which, as we shall see, 
does not fit in the scheme discussed in Section~\ref{sec:2}.
This choice is the one referred to in the
literature \cite{CDE,EKKM2} as the \emph{$ABC$ model}.
The transition rates are the following. If the occupation variables
across the bond $\{x,x+1\}$ are $(\xi,\zeta)$, they are exchanged to
$(\zeta, \xi)$ with rate  $\exp\{V(\xi,\eta)/N\}$
where $V(A,B)=V(B,C)=V(C,A)=-\beta/2$ and  $V(B,A)=V(C,B)=V(A,C)=\beta/2$
for some $\beta >0$.
Therefore the $A$-particles prefer to jump to the left
of the $B$-particles but to the right of the $C$-particles while the
$B$-particles prefer to jump to the the left of the $C$-particles,
i.e.\ the preferred sequence is $ABC$ and its cyclic permutations.
These rates do not satisfy the local detailed balance \eqref{ldb}.

Again by the methods developed in \cite{DPS,KL,KOV}, the hydrodynamic
equations are
\begin{equation}
  \label{hydabc}
  \partial_t
\left(
\begin{array}{c}
\rho_A  \\
\rho_B
\end{array}
\right)
+\nabla \cdot
\left(
\begin{array}{c}
J_A(\rho_A,\rho_B) \\
J_B(\rho_A,\rho_B)
\end{array}
\right)
= 0
\end{equation}
where
\begin{equation}
\label{currABC}
J(\rho_A,\rho_B) =
\left(
\begin{array}{c}
J_A(\rho_A,\rho_B) \\
J_B(\rho_A,\rho_B)
\end{array}
\right)
=
\left(
\begin{array}{c}
 - \nabla \rho_A +\beta \rho_A( 1-2\rho_B - \rho_A) \\
 - \nabla \rho_B +\beta \rho_B(2\rho_A +\rho_B -1)
\end{array}
\right)
\end{equation}
The asymmetric term in the hydrodynamic equation \eqref{hydabc} is not
of the form $\nabla \cdot \big( \chi(\rho) E \big)$ as in
\eqref{2.1}--\eqref{ein_rel}. 
Hence the results obtained in Section~\ref{sec:3} do not apply.

The appropriate cost functional is however still given by \eqref{I=}
with $J$ as in \eqref{currABC} and $\chi$ as in \eqref{chiabc}. The
solution of the Hamilton-Jacobi equation \eqref{HJeq} then gives the
free energy. In the case of equal densities $\int\! dx \, \rho_A
=\int\! dx \, \rho_B = 1/3$, a straightforward computation shows
that for any positive $\beta$ the solution is given by the
functional
\begin{eqnarray}
\label{num}
\nonumber
&&\!\!\!\! \!\!\!
\mc F_{\frac 13,\frac 13}^\beta (\rho_A,\rho_B)
= \mc F^0_{\frac13, \frac 13} (\rho_A,\rho_B)
\\
\nonumber
&& \!\!\!\! 
+ \beta
\int_0^1\!dx \int_0^1 \!dy \: y \, \Big\{
\rho_A(x) \rho_B(x+y) + \rho_B(x) [ 1-\rho_A(x+y) -\rho_B(x+y)]
\\
&& \!\!\!\! \phantom{ \int_0^1\!dx \int_0^1 \!dy \, y \Big\{ \; \; }
+[1-\rho_A(x) -\rho_B(x)] \rho_A(x+y) \Big\} + \varkappa
\end{eqnarray}
where $\mc F^0_{\frac 13,\frac 13}$ is the functional in \eqref{qps=} with
$m_A=m_B=1/3$ and $\varkappa$ is the appropriate normalization constant.
This result has been already obtained in \cite{CDE} by direct
computations from the invariant measure. 
Indeed, in this case, the ABC model is  \emph{microscopically}
reversible and the invariant measure can  be computed explicitly.
From a macroscopic point of view, the reversibility of the model is
expressed as the identity, which holds in the case of equal densities, 
$J(\rho) = J^*(\rho)$. 
Hence, as discussed at the end of Section~\ref{sec:3}, 
the free energy \eqref{num} can also be macroscopically derived by 
integrating \eqref{ein?}.

We mention the recent preprint \cite{BDLW}, of which we became 
aware after the completion of this work, in which the nonequilibrium
free energy of the ABC model is also computed from the
variational principle \eqref{2.6}.

\bigskip\bigskip
\noindent{\bf Acknowledgments.}  
\ We are very grateful to J.L.\ Lebowitz for several stimulating
discussions and to D.\ Mukamel for calling our attention to the ABC
model.  G.J-L.\ is grateful to H.\ Posch and the Schroedinger
Institute (ESI) for the kind hospitality.  L.B. acknowledges the
support of PRIN.  D.G. acknowledges the support of the GNFM Young
Researcher Project ``Statistical mechanics of multicomponent
systems''.

\end{document}